\documentclass[sigconf,natbib=false,authorversion=true,nonacm=true,screen]{acmart}
\pdfoutput=1

\RequirePackage[
  datamodel=acmdatamodel,
  style=acmnumeric,
  ]{biblatex}

\usepackage[toc,acronym,nomain,translate=babel]{glossaries}
\glsdisablehyper
\newacronym{utc}{UTC}{Coordinated Universal Time}
\newacronym{vpn}{VPN}{Virtual Private Network}

\newacronym{ntp}{NTP}{Network Time Protocol}
\newacronym{sntp}{SNTP}{Simple Network Time Protocol}
\newacronym{dns}{DNS}{Domain Name System}
\newacronym{dnssec}{DNSSEC}{DNS Security Extensions}
\newacronym{dhcp}{DHCP}{Dynamic Host Configuration Protocol}
\newacronym{http}{HTTP}{Hypertext Transport Protocol}
\newacronym{https}{HTTPS}{Hypertext Transport Protocol Secure}
\newacronym{upnp}{UPnP}{Universal Plug and Play}
\newacronym{ssdp}{SSDP}{Simple Service Discovery Protocol}
\newacronym{mdns}{mDNS}{Multicast DNS}
\newacronym{igd}{UPnP IGD}{UPnP Internet Gateway Device}
\newacronym{snmp}{SNMP}{Simple Network Management Protocol}
\newacronym{ssh}{SSH}{Secure Shell}
\newacronym{ftp}{FTP}{File Transfer Protocol}
\newacronym{smtp}{SMTP}{Simple Mail Transport Protocol}
\newacronym{tls}{TLS}{Transport Layer Security}
\newacronym{ipp}{IPP}{Internet Printing Protocol}
\newacronym{smb}{SMB}{Server Message Block}
\newacronym{sip}{SIP}{Session Initiation Protocol}
\newacronym{voip}{VoIP}{Voice over IP}
\newacronym{www}{WWW}{World Wide Web}
\newacronym{ssl}{SSL}{Secure Sockets Layer}

\newacronym{nist}{NIST}{National Institute of Standards and Technology}
\newacronym{iana}{IANA}{Internet Assigned Numbers Authority}
\newacronym{rir}{RIR}{Regional Internet Registry}
\newacronym{bgp}{BGP}{Border Gateway Protocol}
\newacronym{isp}{ISP}{Internet Service Provider}
\newacronym{asn}{ASN}{Autonomous System Number}
\newacronym[shortplural={ASes}]{as}{AS}{Autonomous System}
\newacronym{cidr}{CIDR}{Classless Inter-Domain Routing}

\newacronym{xml}{XML}{Extensible Markup Language}
\newacronym{soap}{SOAP}{Simple Object Access Protocol}

\newacronym{ttl}{TTL}{Time-To-Live}
\newacronym{nat}{NAT}{Network Address Translation}
\newacronym{natpmp}{NAT-PMP}{NAT Port Mapping Protocol}
\newacronym{pcp}{PCP}{Port Control Protocol}
\newacronym{turn}{TURN}{Traversal Using Relays around NAT}
\newacronym{cpe}{CPE}{Customer-Premises Equipment}

\newacronym{lan}{LAN}{Local Area Network}
\newacronym{wan}{WAN}{Wide Area Network}

\newacronym{pptp}{PPTP}{Point-to-Point Protocol}

\newacronym{iot}{IoT}{Internet of Things}

\newacronym{ip}{IP}{Internet Protocol}
\newacronym{ipv4}{IPv4}{Internet Protocol version 4}
\newacronym{ipv6}{IPv6}{Internet Protocol version 6}
\newacronym{tcp}{TCP}{Transmission Control Protocol}
\newacronym{udp}{UDP}{User Datagram Protocol}
\newacronym{icmp}{ICMP}{Internet Control Message Protocol}
\newacronym{ipsec}{IPsec}{Internet Protocol Security}

\newacronym{cdn}{CDN}{Content Delivery Network}
\newacronym{vps}{VPS}{Virtual Private Server}
\newacronym{cgn}{CGN}{Carrier-Grade NAT}

\newacronym{hcf}{HCF}{Hop Count Filtering}
\newacronym{dos}{DoS}{Denial-of-Service}
\newacronym{ddos}{DDoS}{Distributed Denial-of-Service}
\newacronym{drdos}{DRDoS}{Distributed Reflective Denial-of-Service}

\newacronym{mpls}{MPLS}{Multiprotocol Label Switching}

\newacronym{ietf}{IETF}{Internet Engineering Task Force}
\newacronym{ripe}{RIPE NCC}{Réseaux IP Européens Network Coordination Centre}
\newacronym{arin}{ARIN}{American Registry for Internet Numbers}

\newacronym{iaas}{IaaS}{Infrastructure as a Service}
\newacronym{paas}{PaaS}{Platform as a Service}
\newacronym{saas}{SaaS}{Software as a Service}

\newacronym{ssrf}{SSRF}{Server-Side Request Forgery}
\newacronym{xxe}{XXE}{XML External Entity}

\newacronym{rfc}{RFC}{Request for Comments}

\glsunsetall

\usepackage{amsmath}
\usepackage{amsfonts}
\usepackage{wasysym}
\usepackage{xargs}
\usepackage{pifont}

\usepackage{titleref}

\newcommand{\citerfc}[1]{{RFC~\citefield{#1}{volume}~\cite{#1}}}
\newcommand{\secref}[1]{{Section~\ref{#1}}}
\newcommand{\figref}[1]{{Figure~\ref{#1}}}
\newcommand{\tabref}[1]{{Table~\ref{#1}}}

\newcommand{\rom}[1]{\textbf{\uppercase\expandafter{\romannumeral #1\relax}}}

\usepackage{xcolor}
\usepackage{booktabs}

\usepackage{caption}
\usepackage{threeparttable}
\usepackage{tabularx}
\usepackage{bytefield}
\usepackage{supertabular}

\begin{document}
\title{\emph{Bad Neighbors}: On Understanding VPN Provider Networks}

\author{Teemu Rytilahti}
\email{teemu.rytilahti@rub.de}
\affiliation{%
  \institution{Ruhr University Bochum}
  \country{Germany}
}

\author{Thorsten Holz}
\email{holz@cispa.de}
\affiliation{%
  \institution{CISPA Helmholtz Center for Information Security}
  \country{Germany}}

\begin{abstract}
    Virtual Private Network (VPN) solutions are used to connect private networks securely over the Internet. Besides their benefits in corporate environments, VPNs are also marketed to privacy-minded users to preserve their privacy, and to bypass geolocation-based content blocking and censorship. This has created a market for turnkey \gls{vpn} services offering a multitude of vantage points all over the world for a monthly price. While \gls{vpn} providers are heavily using privacy and security benefits in their marketing, such claims are generally hard to measure and substantiate. While there exist some studies on the \gls{vpn} ecosystem, all prior works omit a critical part in their analyses: \emph{How well do the providers configure and secure their own network infrastructure?} and \emph{How well are they protecting their customers from other customers?} To answer these questions, we have developed an automated measurement system with which we conduct a large-scale analysis of \gls{vpn} providers and their thousands of VPN endpoints. Considering the fact that VPNs work internally using non-Internet-routable \gls{ip} addresses, they might enable access to otherwise inaccessible networks. If not properly secured, this can inadvertently expose internal networks of these providers, or worse, even other clients connected to their services. 
    Our results indicate a widespread lack of traffic filtering towards internally routable networks on the majority of tested \gls{vpn} service providers, even in cases where no other \gls{vpn} customers were directly exposed. 
    We have disclosed our findings to the affected providers and other stakeholders, and offered guidance to improve the situation. 
\end{abstract}

\settopmatter{printfolios=true}

\maketitle

\section{Introduction}\label{sec:background:badn:turnkey}
The concept of \glspl{vpn} was developed to connect private networks securely over a public network such as the Internet. In practice, \gls{vpn} solutions are often used in corporate environments, e.g., to connect branch offices in different locations. Furthermore, \glspl{vpn} are used by privacy-minded end users to improve their online privacy. Typical use cases in this context include avoiding geolocation-based content blocking and censorship, and preventing the local \gls{isp} from collecting detailed information on the accessed services.

This demand has created a market for turnkey \gls{vpn} services (both free and commercial), which offer various vantage points hosted in different geographical locations around the world. These \gls{vpn} providers market their services claiming security and privacy benefits, which are, however, hard to measure and substantiate in practice~\cite{ramesh__2023}.
For example, in the few past years, several \gls{vpn} providers have been reported to have suffered from security breaches~\cite{markuson_nordvpn_2019,torguard_why_2019,bischoff_zero_2020,vpnmentor_report_2020}.
The sensitive nature of these services requires a better analysis of popular \gls{vpn} providers to assess their security and privacy in more detail.

Previous research in this area has concentrated mainly on understanding potential privacy leaks caused by configuration failures (e.g., routing traffic over the regular connection during connection failures, leaking of \gls{dns} queries by using a different route for queries, etc.~\cite{DBLP:conf/uss/RayAKF12,DBLP:journals/popets/PertaBTHM15}). Furthermore, browser-based information leak sources (like WebRTC~\cite{DBLP:conf/iccst/Al-Fannah17}) to obtain information about the network, and client misconfigurations (e.g., \gls{vpn} applications using insecure settings~\cite{DBLP:conf/cans/ZhangLZWG17,bui_client_side_2019}, or even behaving maliciously~\cite{DBLP:conf/imc/IkramVSKP16}) were examined in prior work. 
Although Khan et al.~\cite{DBLP:conf/imc/KhanDVSKV18} performed a measurement study to understand the intricacies of the \gls{vpn} ecosystem, their analysis left out an analysis on how these providers' \emph{internal} infrastructure is organized and how the providers protect their customers against potential \emph{in-tunnel attackers}, i.e., adversaries who use the same \gls{vpn} service with malicious intents. 

In this work, we fill this gap by presenting an in-depth analysis of insecure configurations of popular \gls{vpn} providers by analyzing their internal infrastructures.
Our study is motivated by several security breaches detected at \gls{vpn} providers. 
For example, in late 2019 three providers (NordVPN, TorGuard, and VikingVPN) were reported to have their servers breached by an adversary, seemingly over an unsecure remote management interface~\cite{bannister_adam_vpn_2019}.
While exact details are unknown, both NordVPN and TorGuard published statements noting that these breaches were isolated incidents on single servers caused by the misconfiguration done by their infrastructure provider~\cite{markuson_nordvpn_2019,torguard_why_2019}.

By covering 67 \gls{vpn} providers and over twenty thousand endpoints using several popular \gls{vpn} protocols (OpenVPN, \gls{pptp}, and WireGuard), we provide an extensive overview of the security posture of \gls{vpn} providers, especially concerning their handling of tunneled network traffic.
\label{badn:sec:threat_model}
More specifically, we observe our findings from the perspectives of three different stakeholders:

\textbf{End Users.}
One of the advertised purposes, besides going around geo\-blocking, of \gls{vpn} services is to improve security and privacy. If a \gls{vpn} leaves a user less safe than without it, it does not fulfill its purpose, and may even expose users to new risks by exposing services that would otherwise be blocked, e.g., consumer \glspl{isp} actively block protocols like \gls{smb} to keep their users safe~\cite{greene_what_2017}.

\textbf{VPN Providers.}
As providers act as a middleman for privacy-conscious users, a breach of a \gls{vpn} server could have more severe implications (e.g., by allowing potential man-in-the-middle attacks) than exposing individual end user systems to other users. Therefore, hardening the \gls{vpn} endpoints should be a priority for service providers especially when they are hosted on shared data centers.

\textbf{Upstreams.}
As the providers are often colocating their servers or renting hosting, incorrectly configured network filtering could potentially put other users using the same facilities in jeopardy. This could be exacerbated by the internal addressing and ``no-log'' policies, making potential incident handling more difficult. 

\begin{figure*}
 \centering
 \includegraphics[width=1.6\columnwidth]{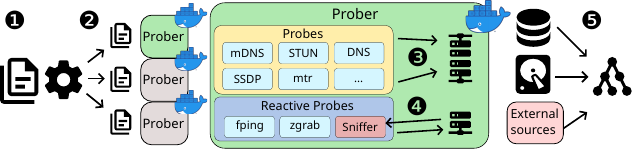}
    \caption[VPN Measurement Approach]{Overview of our scanning setup. \ding{182}~Inspect the provider, register, and obtain configurations. \ding{183}~Generate configurations and launch probing containers. \ding{184}~Start packet capture and execute active probes on connection establishment. \ding{185}~Launch reactive probes when encountering packets from unknown, internal networks. \ding{186}~Combine collected data with in-tunnel packet traces to create a graph. Enrich the graph with information about geolocation, autonomous systems, and Censys information for offline analyses.
    }
 \label{fig:scan_architecture}
 \Description{An overview of our scanning setup, showing different components which are described in the text.}
\end{figure*}

\label{sec:background:vpnprotos}
Given the many different ways in which a \gls{vpn} can be implemented, we constrain ourselves to three widely supported protocols where the testing can be automated: OpenVPN, \gls{pptp}, and WireGuard.
First, OpenVPN~\cite{noauthor_openvpn_nodate} is one of the most commonly offered \gls{vpn} protocols by turnkey \gls{vpn} providers.
First released to the public in 2001, its popularity is likely related to its easy-to-use nature for both client and server implementations and its availability on all common operating systems.
Besides being used on desktop systems, routers, and servers, it is also commonly used as the backend for mobile \gls{vpn} solutions~\cite{DBLP:conf/cans/ZhangLZWG17,DBLP:conf/imc/IkramVSKP16}.

Second, Point-to-Point Tunneling Protocol (PPTP, \citerfc{DBLP:journals/rfc/rfc2637}) is widely popular in practice due to its built-in support on several platforms, most notably Microsoft Windows and Android. 
 While several parts of its implementation are considered to be insecure for a decade (e.g., reported by Schneier~et\,al.~\cite{DBLP:conf/ccs/SchneierM98, DBLP:conf/cqre/SchneierMW99}) and its use has been discouraged by Microsoft---one of its developers---since 2012~\cite{betafred_microsoft_2012}, it is still widely supported.

Third, WireGuard is a fairly new \gls{vpn} protocol that aims to replace \gls{ipsec}, but it is not yet widely supported by \gls{vpn} providers in practice. First introduced by Donenfeld at NDSS~2017~\cite{DBLP:conf/ndss/Donenfeld17}, WireGuard has recently gained more support, e.g., WireGuard has been directly integrated into the Linux kernel since March 2020. 
The lack of a simple option for dynamic addressing has slowed down acceptance among commercial \gls{vpn} providers, although efforts are being made to enable dynamic addressing~\cite{noauthor_wg_dynamic_2021}. 

\smallskip
\noindent
In summary, we make the following three key contributions:

\begin{itemize}
    \item We show in a large-scale measurement study how VPN providers leak sensitive information about their internal infrastructure and expose their clients to in-tunnel attackers. To this end, we perform active measurements on over 20,000 VPN entry points to obtain detailed information regarding the internal infrastructure of these providers with the safety and privacy of their users in mind. We show how deficiencies in the configuration setups leave customers and VPN providers inadvertently exposed to other (potentially malicious) customers. 
    \item Furthermore, we show how routing misconfigurations allow access to services that would otherwise be inaccessible, such as business-critical cloud provider metadata services or inadvertently exposed services at hosting providers' networks, as well as access to other connected but internal networks.
    \item Based on the insights and lessons learned during our study, we provide recommendations and guidance to avoid such problems in the future. Furthermore, we shared our findings with the affected VPN service providers.
    Our disclosure led to configuration changes at several providers.
\end{itemize}

To foster research in this area, we publish several research artifacts under an open-source license at \url{https://github.com/RUB-SysSec/TurnkeyVPNStudy/}.

\section{Approach}
\label{sec:scanner_network_setup}

In this section, we provide a high-level overview of our probing system and the individual components, also depicted in Figure~\ref{fig:scan_architecture}.

\subsection{Obtaining Configuration Information}
\label{sec:configurations}

The first step for testing the security and privacy properties of a given \gls{vpn} provider is the most manually intensive part of our experiments. First, we visit and explore the website of each provider to determine if our system allows testing it.
After this manual review, we obtain access by registering an account with the \gls{vpn} provider. Afterward, we need to locate all relevant configuration information and make them usable for our probing system. This step differs depending on the used protocol.

\textbf{OpenVPN.} Many \gls{vpn} providers provide ready-to-use configuration files for OpenVPN (e.g., in an ideal case even a ZIP file containing all configurations). Analyzing those made it clear that it is common to (i) use domain names in configuration files (e.g., for load balancing), and (ii) host the same service on multiple ports (for getting through network filtering).
As we want to test all potential endpoints but nothing more, we need to accommodate these quirks. 

Based on these insights, we first resolve the domains to their corresponding \gls{ip} addresses. In the second step, we merge all results on the \gls{ip} address per transport protocol.
If the provider offers both \gls{udp} and \gls{tcp} protocol, we ignore the \gls{tcp} configurations.
This approach is based on the assumption that the endpoint does not implement a different routing policy based on the used transport protocol, but is merely using separate configurations for evading filtering.
Empirical tests confirmed this hypothesis.

\textbf{PPTP.}
In contrast to OpenVPN, there is no standard configuration format for PPTP. However, \gls{pptp} configurations are 
much simpler as configurations commonly only differ in single options, e.g., whether MPPE-encryption should be enabled.
We found that providers commonly provide a list of available endpoint addresses and links to some documents that guide users through the setup process.
In many cases, the provider uses the same domains as for OpenVPN, so we simply use the already known addresses for \gls{pptp}, too.
Otherwise, we parse the website for available endpoints.

\textbf{WireGuard.}
WireGuard differs from the two other protocols, as instead of performing authentication with a username/password combination and delivering an \gls{ip} address to use during the connection, it uses \textit{cryptokey routing}~\cite{DBLP:conf/ndss/Donenfeld17} where a configured public key is allowed to use a specific, static \gls{ip} address.
The \gls{vpn} providers we tested allowed either uploading the public key or offered to generate a new key pair that was then used in the generated configuration file.
Every provider we tested provided WireGuard support by reusing the same static, internal \gls{ip} address and the public key on any endpoint, allowing us to generate configurations by simply swapping the endpoints in the configuration.

\subsection{Testing VPN Providers}
\label{sec:testing_a_provider}

We automated the generation of configuration information in a best-effort approach by parsing the configuration files and storing the information in a database for later analysis and generation of connection profiles. To verify that the configuration options are correct, we manually verify that connections are indeed possible using the generated connection profiles.
Once this verification is successful, we can start the actual probing process for each \emph{instance configuration}, i.e., each unique \gls{vpn} configuration we identified. 

\textbf{Concurrent Probing.}
\gls{vpn} providers commonly bestow only a limited number of active connections, making it complicated and time-consuming to test all available instance configurations in an automated way. Our system is hence designed to test concurrently only the number of allowed connections as indicated by the provider while performing probing efficiently.

\textbf{Network Isolation.}
We use Docker~\cite{noauthor_docker_2022} to isolate concurrent \gls{vpn} instances and to make sure we are not accidentally probing our own lab network. To minimize the possibility that our internal docker-to-host communication would clash with the network handed to us by the \gls{vpn} provider, we create unique \texttt{/24} networks for every instance under the reserved \texttt{240/8} network 
(RFC~1112~\cite[``Class E'']{DBLP:journals/rfc/rfc1112} 
for every probing instance testing a single endpoint.

To simplify the data collection process, we mount our host's data volume inside the container (for storing auxiliary output files) as well as expose a database endpoint (i.e., the Unix socket of \emph{MongoDB}~\cite{noauthor_mongodb_nodate}) to allow data storing during the probing.
More details regarding our Docker configuration can be found in Appendix~\ref{sec:appendix:docker}. 

\subsection{Testing an Instance}
\label{sec:testing_an_instance}

The probing system will launch a separate Docker container for each instance configuration. Once a container is started, the first step is to launch our \emph{probing shim}, which will connect to the database to obtain the details of the \gls{vpn} endpoint under test and initialize a VPN connection.
For OpenVPN and \gls{pptp}, we hook the launch of the probing system by defining our shim to be executed on connection establishment.
For WireGuard, we read the return code of the modified \texttt{wg-quick} script.

After a successful connection establishment, the first probing steps are the following:
\begin{enumerate}
 \item Launching the packet capturing process (\textit{tcpdump}) for in-tunnel traffic,
 \item Starting our in-tunnel web server, and
 \item Adjusting the \gls{dns} configuration based on the server-pushed information.
\end{enumerate}

After these steps are finished, the probing process commences with running our tests. 
When all tests are finished, or when the 15-minute time limit is hit, the prober will terminate the VPN process.
Protocol-specific details are provided in Appendix~\ref{sec:protocol_specials}.

\subsection{Ethical Considerations}
Considering that we are studying live systems on the Internet, we designed our experiment to prevent any potential harm on the target systems. While our institution does not require IRB review for minimal risk studies, we ensured that our scanning follows best practices. To this end, we limited our scans to use only a set of probes with benign payloads (i.e., banner grabbing) to avoid causing any kind of disruption. Furthermore, we followed the recommended best practices for Internet scanning introduced by Durumeric~et~al.~\cite{DBLP:conf/uss/DurumericWH13} to make our benign intentions clear.
For HTTP(S) requests, we use a user-agent string containing our contact information and indicating that the requests are made for research purposes.
For SMTP, we use the domain of our scanserver as the domain \texttt{EHLO} greeting.
The server we used for scanning, as well as the in-tunnel probing system, hosts a website explaining our scanning activities and our contact information. The reverse DNS record of the scanning host was set to indicate its use for research purposes. Furthermore, the \emph{whois} information for its IP address contained our abuse e-mail address.
We were not contacted by any party regarding our activities at any time during this measurement study.

With the mentioned precautions in place, we try to balance between the benefits of users and providers alike, and the potential harm caused by our scanning, as discussed in the Menlo report~\cite{DBLP:journals/ieeesp/BaileyDKM12}. We argue that the understanding of these phenomena outweigh the unlikely potential harms, especially considering the fact that these services are in-fact marketed with their security and privacy benefits (e.g., with ``no-log'' policies) which may leave providers themselves unknowledgeable about potential issues. 

\textbf{Responsible disclosure.}
We reached out to the affected 61 \gls{vpn} providers and informed them about our findings, with mostly positive responses.
In total, we received responses from 31 providers which 30~\% were non-automated, and finally a handful of vendors reported configuration changes thanks to our disclosure.
We provide more details on the disclosure process in \secref{sec:appendix:disclosure}.

\section{Probes}\label{sec:probe_details}
Next, we describe the different types of probes used by our analysis system.
Given the scale of our measurements, it is important to optimize the run-time for probing an instance to efficiently test as many endpoints as possible. The problem we try to solve is two-fold: 
\begin{enumerate}
\item In contrast to Internet-wide network scanning, we do not know the address space that we are about to scan, and more importantly, 
\item The number of simultaneously active connections is limited by the \gls{vpn} provider which may have thousands of endpoints, so we want to optimize how much time is spent on testing a single endpoint.
\end{enumerate}

To optimize our scanning process while avoiding missing interesting data points,
we categorize the probes into two categories:

\begin{itemize} 
 \item Active probes are launched immediately after the connection establishment and are responsible for actively probing the network we are connected to. These include traceroutes to selected targets, running the \gls{ip} spoofing tests, and many other probes.
 \item Reactive probes will act on input from other probes in a publish-subscribe setting, most prominently to perform application layer scans using ZGrab2~\cite{durumeric_search_2015}.
\end{itemize}

As soon as all probes in the first category have finished their tasks, the probing system will not forward any new tasks to the reactive probes.
Instead, it will wait for existing tasks to finish, or the timeout to hit, at which point the connection will be disconnected.

\textbf{Packet sniffer.}
No matter how carefully we construct our probes and their result handling, we can never obtain a full picture of what is transmitted on the wire.
Therefore, we listen for incoming packets and add all encountered internal networks to the work queue for reactive probes.
For example, we use CAIDA's \emph{IP spoofer} program~\cite{beverly_spoofer_2005} to check if \gls{ip} spoofing is possible, potentially provoking unexpected responses from hosts we were not explicitly targeting.

\subsection{Active Probes}
\label{sec:proactive_probes}
Probes in this category attempt to collect information about the network topology and hosts within the network.
While performing these tests, a secondary goal is to elicit responses from as many internal hosts as possible to locate accessible hosts.
All active probes are launched and run concurrently to collect information and to fill the work queue for reactive probes, and they are always executed until they are completed. We use the following probes:

\textbf{Environment Exploration.}
For some \gls{vpn} providers, some information is already available directly in the configuration files, e.g., WireGuard configurations include the local \gls{ip} address as well as the \gls{dns} servers to use.
For OpenVPN and \gls{pptp}, this information is pushed by the server during the connection initialization phase.

\textbf{Vantage Point Information.}
As soon as the connection has been established, we use an external service (ipify~\cite{noauthor_ipify_nodate}) to find out the \gls{ip} address of the vantage point (i.e., egress) to perform reverse \gls{dns}, geolocation, and \gls{as} lookups.
In this step, we also perform STUN checks (using pystun~\cite{pystun}) to determine what type of \gls{nat} is in use. 

\textbf{Traceroutes.}
We perform several traceroutes using mtr~\cite{cross_official_2021} to map the network. 
The intuition is to elicit responses from hosts along the path to expand the set of accessible private networks.
To this end, we target the following \gls{ip} addresses using both \gls{icmp} and \gls{tcp} modes:
\begin{enumerate}
\item First and last host address of all RFC~1918 ranges,
\item A server under our control as an external host,
\item 8.8.8.8 (Google's DNS), and
\item The \emph{link-local} address 169.254.169.254 (RFC~3927~\cite{DBLP:journals/rfc/rfc3927}).
\end{enumerate}

\textbf{Link-local Service Discovery.}
Protocols such as \gls{dns} Service Discovery (DNS-SD) over \gls{mdns} and \gls{ssdp} are used to locate link-local devices in home networks.
We send probes using these protocols, as they allow discovering devices over multicast without actually knowing their addresses.

\textbf{Cloud Metadata Services.}
Many cloud providers implement a service to allow requesting metadata (and user-stored data, like API tokens) over a common location~\cite{amazon_retrieve_nodate,kumarisupriya_azure_nodate,google_storing_nodate}.
Such a service is commonly hosted at the link-local \gls{ip} address \texttt{169.254.169.254}.
Unauthorized access to such services can lead to security issues,  as was seen in an incident at the Capital One bank in 2019~\cite{noauthor_2019_2019}.
To verify that such accesses are not allowed, we try to access the metadata services. 

\textbf{IP Spoofing.}
The presence of networks that do not perform source filtering is a known root cause for \gls{ddos} attacks~\cite{AmplificationHell,DBLP:conf/uss/KuhrerHRH14}.
We use a tool from CAIDA's \emph{IP spoofer} tool~\cite{beverly_spoofer_2005} to test if the endpoint filters incoming and outgoing spoofed traffic.
As a side effect, this probe \emph{may} elicit responses from unexpected networks.

\subsection{Reactive Probes}\label{sec:reactive_probes}

In contrast to active probes, reactive probes listen in the background and wait for \gls{ip} addresses and networks to arrive in the work queue.
Our architecture follows a publish-subscribe model, where the reactive probes wait for other probes (which can also be reactive themselves) to provide networks to process.
When encountering a non-Internet-routable \gls{ip} address,
we add its \texttt{/24} subnet and its adjacent subnets to our list of networks to probe.
In case this network has not already been processed, the
handlers for all reactive probes get called with this new network.
Next we describe two types of reactive probes: ping sweeps and ZGrab2 probes.
The intuition for doing \emph{ping sweeps} (using fping~\cite{schweikert_fping_nodate}) on found networks comes from the fact that networks from which we have seen responsive hosts may also have other accessible hosts.

As having received a packet does not provide much information in itself, 
we use an extended version of the ZGrab2~\cite{durumeric_search_2015} network scanner to run application-layer scans to quantify our findings.
We scan every network using a selected set of built-in protocols (HTTP(S), SMTP, FTP, SSH, Telnet, SMB, NTP, and IPP) and our additions SNMP, NetBIOS, DNS, and UPnP.
Table~\ref{tab:zgrab_probes} shows the used ZGrab2 probes, including the request details and whether we consider them to have strong identifiers (discussed later in \secref{sec:graph}).

\begin{table}
	\centering
	\caption{Used ZGrab2 Probes}
\begin{threeparttable}
\begin{tabular}{lrllc}
\toprule
              Probe &  Port & Protocol &                 Request &  Strong \\
\midrule
               HTTP &    80 &      TCP & GET request $\dagger$ & \Circle \\
              HTTPS &   443 &      TCP & GET request $\dagger$ & \CIRCLE \\
               SMTP &    25 &      TCP &   send-ehlo $\dagger$ & \Circle \\
                FTP &    21 &      TCP &                     - & \Circle \\
                SSH &    22 &      TCP &                     - & \CIRCLE \\
             Telnet &    23 &      TCP &                     - & \Circle \\
                SMB &   445 &      TCP &                     - & \Circle \\
    DNS $\ddagger$  &    53 &      UDP &        “version.bind” & \Circle \\
                NTP &   123 &      UDP &                     - & \Circle \\
                IPP &   631 &      TCP &                     - & \Circle \\
NetBIOS $\ddagger$  &   139 &      UDP &              NBSTAT * & \Circle \\
  SNMPv2 $\ddagger$ &   161 &      UDP &        GET BULK MIB-2 & \Circle \\
  SNMPv3 $\ddagger$ &   161 &      UDP &        GET BULK MIB-2 & \CIRCLE \\
    UPnP $\ddagger$ &  1900 &  UDP/TCP &                     - & \CIRCLE \\
\bottomrule
\end{tabular}

 \begin{tablenotes}
  \item[$\dagger$] Indicator for benign use: user-agent, ehlo-domain.
  \item[$\ddagger$] Our ZGrab2 extensions.

  \end{tablenotes}
  \end{threeparttable}
\label{tab:zgrab_probes}
\end{table}

\textbf{SNMP.}
\gls{snmp} is the industry standard for monitoring network devices and is expected to be found on any maintained network.
We use gosnmp~\cite{gosnmp} library to query \gls{snmp} information using both widely used protocol versions.
We perform a \texttt{GET BULK} request 
on the \textit{MIB-II} (1.3.6.1.2.1.1)~\cite{DBLP:journals/rfc/rfc1213} containing information like the system name.
We use the commonly known default ``public'' community~\cite{cert_advisory_cert_2002,the_shadowserver_foundation_shadowserver_nodate}.
Version 3 is more complex and requires authentication, 
however, the protocol is noisy also for non-authenticated requests.

\textbf{NetBIOS.}
NetBIOS~\cite{DBLP:journals/rfc/rfc1001, DBLP:journals/rfc/rfc1002} is used in conjunction with \gls{smb} to provide device and service discovery and hostname resolution on Windows systems.
We use the wildcard query (\texttt{NBSTAT *}) 
to return the list of all known hosts, services, and workgroups.

\textbf{DNS.}
Home routers commonly host a caching \gls{dns} resolver, such as dnsmasq, for their clients.
We perform a single \texttt{version.bind} query to detect such services.

\textbf{UPnP.}
UPnP and its service discovery protocol (\gls{ssdp}) are widely used in home networks.
We send a wildcard query (\texttt{ssdp:all}) for \gls{udp} unicast discovery~\cite{natscan_2020}.

\section{Experimental Setup}\label{sec:badn:experimental_setup}

We used \textit{That One Privacy Site}'s \gls{vpn} listing~\cite{thatoneprivacysite} as our primary source for VPN providers, given that it contains informational entries (including information about pricing, supported protocols, and more) for a total of 122 providers.
We extend this list with the providers tested by Khan~et~al.~\cite{DBLP:conf/imc/KhanDVSKV18} and manually add several others, summing to a total of 173 providers.
We manually investigated the websites of these providers to determine which we could evaluate.
In total, over half of the providers were excluded from our study. Common exclusion reasons include:
\begin{itemize}
\item Not supporting any protocols our probing system supports, 
\item Providers not existing anymore (e.g., due to mergers or discontinued businesses),
\item Not being able to register or make a payment to them, or
\item Unreasonable trial or refund conditions (e.g., meager data allowances for trial or lengthy minimum subscriptions of several months length).
\end{itemize}
In the end, we tested 67 providers. This list includes 39 (63~\%) providers from Khan~et~al.'s~\cite{DBLP:conf/imc/KhanDVSKV18} list (most not tested were either discontinued, did not accept our payments, or did not support required protocols), and we added further 28 providers that they did not test.
A complete list of tested providers---including the used payment method---can be found in Table~\ref{tab:tested_providers} in Appendix~\ref{sec:appendix:providers}.

\subsection{Obtaining Access}

For registration, we used a separate e-mail address for every tested service in the form of \texttt{<provider>@\-<ourdomain>.xyz}.
The reason for this is two-fold:
\begin{enumerate}
\item To avoid giving out our personal e-mail accounts on potentially ``shady'' providers (e.g., \emph{Safe-Inet} was taken down pre-disclosure in a Europol-coordinated operation~\cite{europol_cybercriminals_2020}), and 
\item To make it easy to separate communication with providers, if necessary.
\end{enumerate}
We configured the domain at the registrar to forward all e-mails to us.
Furthermore, we pointed the domain's A record to our server hosting the website describing our research intention.

We performed our tests iteratively, one provider at a time, roughly in the following order:
\begin{enumerate}
\item Providers with free trials,
\item Providers that accepted our payments using prepaid credit cards,
\item Providers accepting cryptocurrencies, and lastly
\item We used PayPal or our own credit card for the remaining providers listed by Khan~et~al.~\cite{DBLP:conf/imc/KhanDVSKV18} to make our \gls{vpn} selection comparable to theirs.
\end{enumerate}

Out of 67 tested providers, 24 (36~\%) were either completely free or offered a free trial.
12 providers (18~\%) accepted our prepaid credit cards, and 17 (25~\%) accepted Bitcoins.
As the last resort to test some popular providers, for 11 (16~\%) providers, we had to use either PayPal or our own credit card.
Finally, we got access to two providers using vouchers handed to us by third parties.
After refunds, we spent around 200 USD to perform our initial study.

\subsection{Data Collection \& Analysis}\label{sec:graph}

To perform our analyses, we construct a graph database from the data saved directly to a MongoDB database during the probing phase, as well as the information stored in auxiliary files (e.g., PCAP files and ZGrab2 result files)
We ignore all instances where the packet trace contains packets sourced or destined from our network to make sure that we are not scanning our own internal networks.
In our graph, each \textit{Provider} node connects to an \textit{Instance} for each tested configuration that can expose multiple \textit{Services}.
Each service node has a unique identifier that is constructed from the response payload (detailed later in Section~\ref{sec:protocol_analyses}).
We create an edge for each individual exposure
and the endpoint where it was encountered.

We parse the PCAPs to enrich our graph by adding \gls{ttl} and hop count information to the edges between endpoints and exposed services.
While doing that, we ignored RST-flagged \gls{tcp} packets as such responses were encountered from middleboxes on some providers.
As the initial \gls{ttl} values vary among operating systems,
we convert the values to estimated hop count values by using the lowest nearest initial \gls{ttl} from the following candidates: 30 (routers), 64 (e.\,g.~Linux), 128 (Windows), and 255 (routers).

In contrast to Internet-wide scans, where \gls{ip} addresses can be used as unique identifiers, our study requires a different approach.
For example, an individual \gls{ssh} server may be accessible on multiple endpoints in different addresses of the same \gls{vpn} provider.
To estimate the number of uniquely exposed services, we hence leverage protocol-specific identifiers: 

\begin{itemize}
\item From the tested protocols (cf., Table~\ref{tab:zgrab_probes}), we consider \gls{ssh}, \gls{https}, \gls{snmp}v3, and \gls{upnp} to provide \textit{strong identifiers}.
While the protocols themselves do not guarantee that a given service is necessarily unique (e.g., \gls{tls} certificates and \gls{ssh} host keys could be shared), we consider them as a proxy for uniqueness.

\item To facilitate the analyses for protocols without specified or standardized unique identifiers (\textit{weak identifiers}), we construct a \textit{pseudo identifier} (i.e., an SHA-256 hash) over selected parts of the response payloads.
\end{itemize}

\section{Evaluation}\label{sec:badn:eval}

\begin{table}
\small
    \centering
    \caption{Summary of Tested Providers and Endpoints.}
\begin{threeparttable}
\begin{tabular}{lrrrrr}
\toprule
{} & Providers & \%~$\dagger$ & Endpoints & \%~$\dagger$ & Exposes~$\ddagger$ \\
\midrule
OpenVPN   &        65 &         90.8 &    15,566 &         49.7 &             21,976 \\
PPTP      &        31 &         83.9 &     3,744 &         78.0 &              6,611 \\
WireGuard &         9 &         77.8 &       798 &         26.3 &              1,998 \\
\midrule
Total     &        67 &         91.0 &    20,108 &         54.0 &             23,518 \\
\bottomrule
\end{tabular}

 \begin{tablenotes}
 \item[$\dagger$] Percentage of Exposing $\ddagger$ Exposed Unique Identifiers
 \end{tablenotes}
  \end{threeparttable}
\label{tab:protocol_stats}
\end{table}

Overall, we tested a total of 67 \gls{vpn} providers with over 20,000 endpoints.
All but two providers supported OpenVPN, while nine were offering WireGuard support. Surprisingly, almost half of the providers still offered support for \gls{pptp}, which has been considered insecure for over a decade~\cite{DBLP:conf/ccs/SchneierM98, DBLP:conf/cqre/SchneierMW99} and even Microsoft has dissuaded its use since 2012~\cite{betafred_microsoft_2012}.
A summary of the tested providers and the distribution among protocols is shown in Table~\ref{tab:protocol_stats}.

We ran the experiments iteratively one provider at a time over the summer and fall of 2020. We collected roughly 2 terabytes of raw data, most of it in-tunnel network traces.
Based on the packet traces---excluding a slight overhead caused by 
connection establishments---testing a single endpoint took on average four minutes with 97~\% of all endpoints tested in less than five minutes.
Only in a few cases was the hard time limit of 15 minutes reached.

We focus our analyses on the ZGrab2 results, as reporting merely on received packets does not give a clear picture of the exposed hosts: for some providers, \gls{smtp} and \gls{dns} requests were responded to by (seemingly) every host.
To put this in numbers, the average number of responsive internal hosts for an endpoint was over 30,000 ($\sigma$=\textasciitilde80,000) on over 300 different internal /24 networks ($\sigma$=\textasciitilde800), median being over 4,000 hosts on over 100 networks.

In total, we received 270,000 unexpected responses from almost 11,000 \gls{vpn} endpoints on 61 providers.
If not stated otherwise, we will use the unique identifiers instead of individual exposures in further analyses.
While this will cause us to underreport the exposed services---especially on protocols with weak identifiers---this gives us a clear lower bound of the exposed services.
The exposed services had about 24,000 unique identifiers and were seen on endpoints in 79 different countries and in 428 different autonomous systems (based on the endpoint's \gls{ip} address).

On average \textit{for all endpoints}, an endpoint exposed 5 services (median 1, $\sigma$=43) showing extreme skew towards fewer endpoints exposing most of the identifiers.
Table~\ref{tab:exposed_stats} provides a summary of the exposed services per protocol which we will discuss later per protocol.
Note that the vast majority of these exposed services---aside a few exceptions, e.g., providers offering in-tunnel \gls{http} or \gls{ssh} proxying---should not have been accessible at all.

\begin{table*}
\small
 \centering
 \caption{Exposed Services per Protocol. Filtering described in Section~\ref{sec:censys_filtering}. The percentage columns show the proportion of exposures on given protocol. Identifiers show the number of unique identifiers, and Exposes the number of individual exposures.}
 \begin{threeparttable}
 \resizebox{1.0\columnwidth}{!}{\begin{tabular}{lrrrrrr}
\toprule
\multicolumn{7}{c}{Unfiltered} \\
\midrule
\multicolumn{1}{l}{Protocol} & \multicolumn{2}{c}{Providers} & \multicolumn{2}{c}{Endpoints} & \multicolumn{2}{c}{Exposures} \\
\cmidrule(lr){2-3} \cmidrule(lr){4-5} \cmidrule(lr){6-7}
{} & Count & \% &  Count & \% & Identifiers & Exposes~$\dagger$ \\
\midrule
SSH          &        57 &     85.1 &     3,596 &     17.9 &              3,829 &            17,101 \\
HTTPS        &        54 &     80.6 &     1,497 &      7.4 &              1,464 &            15,287 \\
SNMPv3       &        53 &     79.1 &     4,872 &     24.2 &              9,145 &           173,684 \\
UPnP         &        30 &     44.8 &       268 &      1.3 &                222 &               405 \\
\midrule
Strong ident &        60 &     89.6 &     7,561 &     37.6 &             14,660 &           206,477 \\
\midrule

HTTP         &        58 &     86.6 &     4,966 &     24.7 &              4,336 &            38,833 \\

Telnet       &        49 &     73.1 &       622 &      3.1 &                413 &             8,428 \\
FTP          &        42 &     62.7 &       470 &      2.3 &                595 &             6,018 \\
NetBIOS      &        39 &     58.2 &     1,241 &      6.2 &              1,060 &             2,185 \\
SNMPv2       &        35 &     52.2 &       335 &      1.7 &                846 &             8,209 \\
SMB          &        33 &     49.3 &       827 &      4.1 &              1,209 &             2,052 \\

IPP          &        15 &     22.4 &        90 &      0.4 &                 17 &               113 \\
SMTP         &        13 &     19.4 &        97 &      0.5 &                382 &             1,250 \\
\midrule
Weak ident   &        59 &     88.1 &     6,143 &     30.6 &              8,858 &            67,088 \\
\midrule
Total        &        61 &     91.0 &    10,861 &     54.0 &             23,518 &           273,565 \\
\bottomrule
\end{tabular}
}
 \quad
 \resizebox{1.0\columnwidth}{!}{\begin{tabular}{lrrrrrr}
\toprule
\multicolumn{7}{c}{Filtered} \\
\midrule
\multicolumn{1}{l}{Protocol} & \multicolumn{2}{c}{Providers} & \multicolumn{2}{c}{Endpoints} & \multicolumn{2}{c}{Exposures} \\
\cmidrule(lr){2-3} \cmidrule(lr){4-5} \cmidrule(lr){6-7}
{} & Count & \% &  Count & \% & Identifiers & Exposes~$\dagger$ \\
\midrule
SSH          &        56 &     83.6 &     1,233 &      5.9 &              2,146 &             6,667 \\
HTTPS        &        46 &     68.7 &       377 &      1.8 &                685 &             2,876 \\
SNMPv3       &        50 &     74.6 &     1,178 &      5.6 &              1,413 &            18,692 \\
UPnP         &        25 &     37.3 &       127 &      0.6 &                116 &               190 \\
\midrule
Strong ident &        58 &     86.6 &     1,841 &      8.8 &              4,360 &            28,425 \\
\midrule

HTTP         &        49 &     73.1 &       951 &      4.5 &              1,175 &            12,707 \\
Telnet       &        49 &     73.1 &       622 &      3.0 &                410 &             8,411 \\
FTP          &        31 &     46.3 &       228 &      1.1 &                131 &             1,112 \\
NetBIOS      &        34 &     50.7 &     1,153 &      5.5 &                632 &             1,718 \\
SNMPv2       &        33 &     49.3 &       275 &      1.3 &                784 &             7,218 \\

SMB          &        33 &     49.3 &       826 &      4.0 &              1,186 &             1,952 \\

IPP          &        15 &     22.4 &        89 &      0.4 &                 15 &               111 \\
SMTP         &         9 &     13.4 &        40 &      0.2 &                311 &               316 \\
\midrule
Weak ident   &        54 &     80.6 &     2,343 &     11.2 &              4,644 &            33,545 \\
\midrule
Total        &        59 &     88.1 &     2,982 &     14.3 &              9,004 &            61,970 \\
\bottomrule
\end{tabular}
}
 \begin{tablenotes}
 \item[$\dagger$] A unique identifier can be exposed multiple times

\end{tablenotes}
\end{threeparttable}
\label{tab:exposed_stats}
\label{tab:non_censys_stats}
\end{table*}

\subsection{Filtering Out Internet-visible Services}\label{sec:censys_filtering}
As we are mostly interested in hosts that are only
available in-tunnel, we use Censys' BigQuery database~\cite{durumeric_search_2015} to filter out the services that are also visible to the Internet.
We use a single date (2020-09-01) within our testing period to provide an estimation of non-Internet exposed instances, except for \gls{https} where we compared the identifiers to their complete certificate collection.
For protocols that we were not able to get information from Censys (\gls{snmp}v3, NetBIOS, \gls{upnp}, \gls{ipp}), we consider the identifiers not accessible over the Internet if any other service hosted on the same host is considered as such.
We will discuss the constructed protocol-specific identifiers later in Section~\ref{sec:protocol_analyses}.

The right-hand side of Table~\ref{tab:exposed_stats} gives a summary of the dataset after filtering, which we will use for the rest of the analyses.
Although the filtering more than halves the seen unique identifiers--caused largely by the decrease in \gls{snmp}v3 (where we have no ground truth) and \gls{http} (where we received responses from proxies) identifiers---we note that the proportion of exposing providers remains similar.

\subsection{Analyses Overview}\label{sec:results_overview}
Before going to analyze protocol-specific details,
we begin with an overview of the services from the perspective of all individual stakeholders described in Section~\ref{badn:sec:threat_model}.
Given that there is no clear-cut way---due to different ways networks can be configured---to decide whether exposed services belong to an end user in contrast to the \gls{vpn} provider infrastructure or to their upstreams, we use the combination of hop count information, network difference (i.e., is the exposure in the same \texttt{/24} subnet) and selected protocols to construct plausible heuristics with which we can categorize our results and gain some insights.

\begin{figure}
  \centering
  \includegraphics[width=.95\columnwidth]{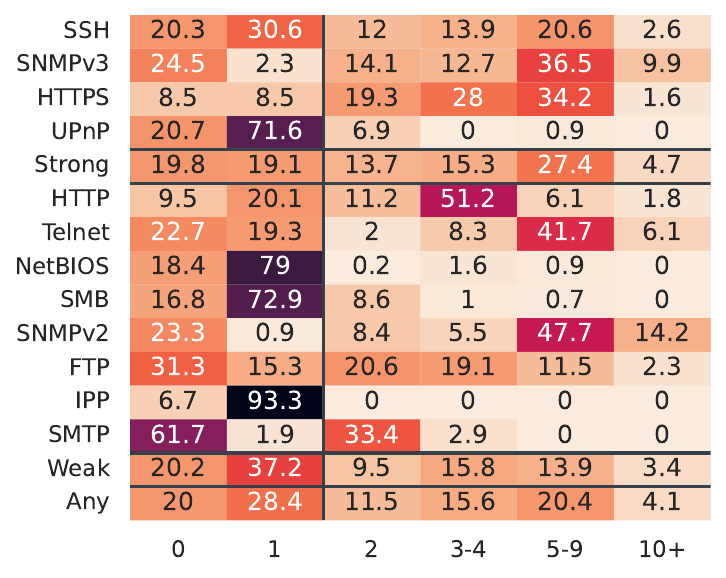}
  \caption[Distribution of Exposed Services per Hop Distance]{Distribution of Exposed Services per Hop Distance. The proportion of protocols commonly found on end-user systems (\gls{upnp}, NetBIOS, \gls{smb}, \gls{ipp}) are more often closer in hops than the protocols used for administrative uses.}
  \label{fig:hop_distance}
  \Description{A heatmap showing the distribution of exposed protocols on different hop count distances. Shows how the proportion of protocols commonly found on end-user systems (\gls{upnp}, NetBIOS, \gls{smb}, \gls{ipp}) are more often closer in hops than the protocols used for administrative uses.}
\end{figure}

\textbf{Hop Distances per Protocol.}
First, Figure~\ref{fig:hop_distance} shows the proportion of exposed services per protocol for different distances measured by hop counts.
We show zero, one, and two-hop results separately to highlight the changes in exposed protocols among distances.
In cases where the same identifier was exposed at different distances, we use here the smallest distance.

In summary, half of all unique exposed services were either zero hops (20~\%) or one hop (28~\%) away from our probing system.
As can be seen in the figure, many, especially end-user protocols were overwhelmingly not directly connected but a single hop away.
Without knowing the exact network structure of the providers, one potential reason for this one-hop deviation could be caused by different routing setups. 
Case in point, OpenVPN's internal routing does not go through the network stack responsible for decrementing the \gls{ttl}.
The other larger cluster of exposures was surprisingly rather farther away---concentrating especially between 5 to 9 hops---hinting towards services hosted by upstream providers.

\begin{figure*}
  \centering
\includegraphics[width=2.0\columnwidth]{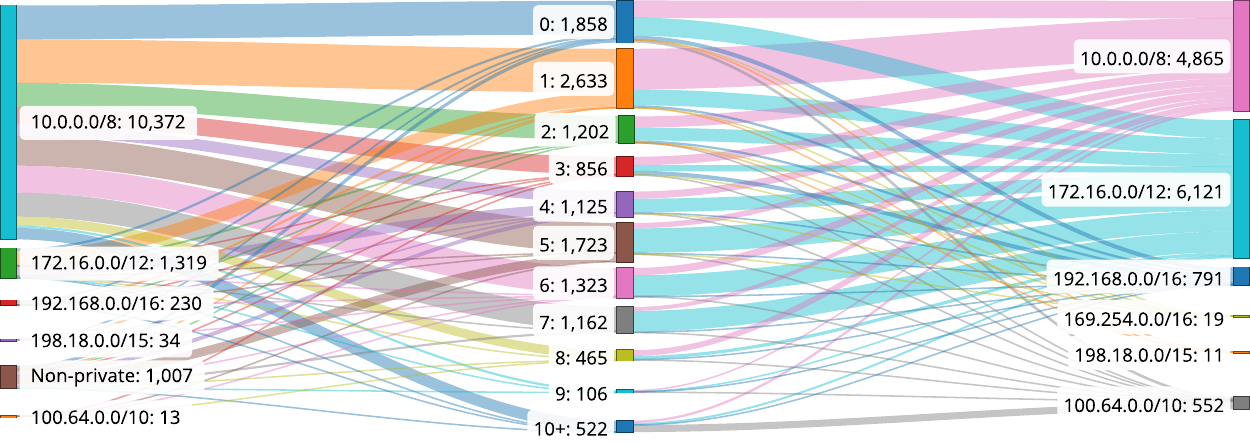}
  \caption[Source and Destination Networks per Hop Count]{Source and Destination Networks per Hop Count. The left-hand side shows the network given to our client whereas the right-hand side shows the network where the exposed service was seen.}
  \label{fig:sankey_src_hops_dst_services}
  \Description{A sankey diagram that shows how the exposures are distributed among hop counts, and source and destination networks. This shows how most of the exposures were commonly located in a different network than which was assigned to the client.}
\end{figure*}

\textbf{Source and Destination Networks.}
Secondly, we take a look at where the exposures are located to our probing system in terms of addressing and hop count distances.
To this end, we group the unique identifiers to the aggregated, RFC 1918 networks on both sides of the connection.
The left side in \figref{fig:sankey_src_hops_dst_services} shows the network where our prober was located and the right side where the exposed service was located, while the number of hops between the prober and the exposed service is shown in the middle.
This highlights how exposures with lower hop counts were often in the same network, while showing that hop count alone is not sufficient for categorization.
Note that these numbers do not sum up with the ones in Table~\ref{tab:exposed_stats} as the identifier may have been seen in multiple networks on different providers.

We emphasize that while \texttt{10/8}---OpenVPN's default network---was the network in which our prober got its address assigned for most of the exposures, most of the exposures themselves were actually located in other networks, predominantly in the \texttt{172.16/12} network.
This \emph{strongly} implies that these exposures are not end-user exposures---assuming other users would be handed out addresses from the same address space as we received ours---but rather exposed services belonging to other stakeholders.
This becomes even more clear when looking at the discrepancy in proportions among hops: a majority of exposures with a hop distance of more than two are predominantly outside the client-given address space range.

We were also assigned addresses from unexpected networks: \texttt{198.18.0.0/15} (Benchmarking, \citerfc{DBLP:journals/rfc/rfc2544}) and \texttt{100.64.0.0/10} ( 	Shared Address Space, CGNAT, \citerfc{DBLP:journals/rfc/rfc6598}).
Six providers also assigned addresses from public address spaces, with two providers (totaling over 500 unique identifiers) squatting the address space of the UK's Ministry of Defence (\texttt{25/8}).
Unsurprisingly, the popular \texttt{192.168.0.0/16} network---commonly used by home routers---was not that common,
likely to avoid routing this address space over the tunnel to allow users to continue to access the devices that are directly connected to their home networks.

\begin{figure}
  \centering
  \includegraphics[width=1\columnwidth]{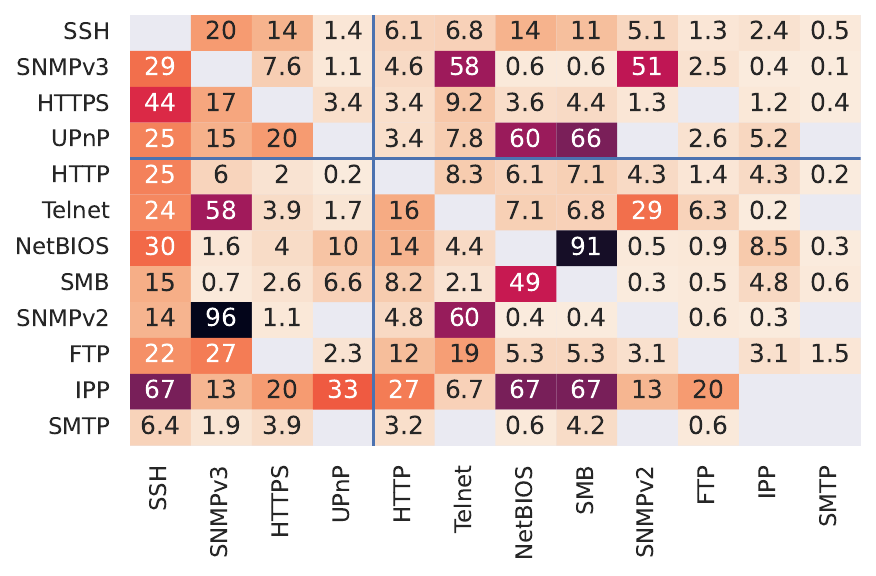}
  \caption[Proportion of Hosts with Shared Identifiers]{Proportion of Hosts with Shared Identifiers. This shows the probability of how often the protocol on the X axis was seen on the same host as the protocol on Y axis. For example, 91~\% of hosts exposing NetBIOS did also expose \gls{smb}, while the case was only 49~\% the other way around.}
  \label{fig:protocols_seen_together} 
  \Description{A heatmap showing the proportion of hosts with shared identifiers. This shows the probability of how often the protocol on the X axis was seen on the same host as the protocol on Y axis. For example, 91~\% of hosts exposing NetBIOS did also expose \gls{smb}, while the case was only 49~\% the other way around.}
\end{figure}

\textbf{Co-hosted Services.}
Thirdly, it is interesting to see which services are exposed together.
To that end, Figure~\ref{fig:protocols_seen_together} shows the proportion of how often a given protocol (on the X axis) was seen with another one (on the Y axis) on the same host.
For example, when \gls{https} service was being exposed, there was 44~\% chance that the same host was also offering \gls{ssh}.
If a host exposed \gls{snmp}v2, it is almost certain that \gls{snmp}v3 was also exposed, and more than half of the time Telnet was also exposed.

This shows us that there are protocols that are more commonly seen together.
On the one hand, especially \gls{upnp}, NetBIOS, \gls{smb}, and \gls{ipp}---protocols commonly found on end-user devices---are commonly seen together.
On the other hand, both \gls{snmp} versions go oftentimes hand in hand with Telnet while \gls{ssh} is seen often with several of the protocols.

\textbf{Autonomous Systems and Locations.}
Lastly, we take a peek at the autonomous systems and locations where we encountered exposures to see if the problem is focused on a very specific set of endpoints.
To do that, we used pyasn~\cite{asgharipyasn} to obtain the \gls{as} information and MaxMind's GeoIP Lite2~\cite{GeoIP} to determine the geolocation for the vantage points (i.e., egress \gls{ip} addresses).
Overall, we found exposed services on 291 different autonomous systems and in 72 different countries,
highlighting that our findings are neither limited to a handful of networks nor countries.
In Appendix~\ref{sec:appendix:vantagepoints}, \tabref{tab:instance_asn} shows the most common ASes and \tabref{tab:instance_country} the most common countries with exposed services,
showing that some networks were disproportionally overrepresented in our findings.
This clearly shows that there are differences in the way upstream hosting services perform network-level filtering and highlights the need for careful testing when it comes to understanding the ecosystem as a whole.

\subsection{Stakeholder Analysis}\label{sec:stakeholders}
We now discuss how we use the preliminary analyses to categorize the exposures to the stakeholder groups.
To do that, we leverage the information shown in \figref{fig:hop_distance}, and \figref{fig:protocols_seen_together}, and \figref{fig:sankey_src_hops_dst_services}, to build our categorization heuristics.

A strict, but straightforward way to classify everything on the same network link (i.e., with $hops = 0$) as end users sounds lucrative, but a look at \figref{fig:hop_distance} shows that protocols commonly found on end-user systems---NetBIOS, \gls{smb}, \gls{upnp}, and \gls{ipp}---were surprisingly often not directly connected.
These protocols were also more often seen together as shown in \figref{fig:protocols_seen_together}.
Likewise, many of the protocols rarely used on end-user devices, like \gls{ssh} and \gls{snmp}, were often seen close to our measurement point.
When looking at other protocols seen in combination with the commonly seen \gls{ssh} on zero and one-hop distances, 
many of the response payloads hinted towards consumer devices.

\begin{figure*}
  \centering
\includegraphics[width=2.0\columnwidth]{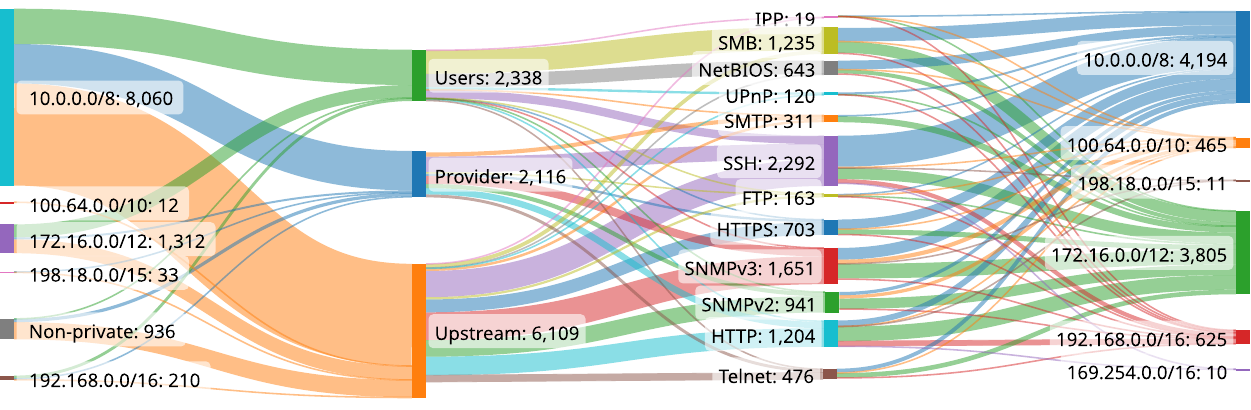}
  \caption[Overview of Stakeholder Categorized Exposures]{Overview of Stakeholder Categorized Exposures. Left: prober's network. Right: exposed service's network.}
  \label{fig:sankey_src_category_protocol_dst}
  \Description{A sankey showing an overview of exposures by stakeholder, protocol, and source \& target networks.}
\end{figure*}

Trying to categorize services to belong to the provider instead of an end user is more complex.
While services like \gls{snmp}, \gls{ftp}, \gls{smtp}, Telnet, and \gls{ssh} are uncommon on end-user systems, Linux-based home routers are common and many \gls{vpn} providers even give instructions for setting their service on a home router.
The services that are deeper into the network are likely from the upstream providers of the \gls{vpn} service.
Looking at the portions of exposed protocols around the clusters between 3-9 hops makes it quite obvious that these likely belong to some upstream providers.
The \gls{snmp} information, \gls{https} certificates with subjects related to common network gear, and Telnet banners indicate at times also names of their owner.
With these considerations and based on our exploratory data analyses, we chose the following heuristics:
\begin{enumerate}
    \item \textbf{End Users}: Every end-user protocol (NetBIOS, \gls{smb}, \gls{upnp}, and \gls{ipp}) exposure (i) with a maximum hop count of 1, or (ii) where both the exposing host and our prober are in the same \texttt{/24} network.
    If a host in this category exposes other services, we also mark them as end-user exposures.
    \item \textbf{VPN Providers}: Exposures with characteristics above \emph{without} any end-user protocols.
    This may cause us to underestimate the provider infrastructure as not every end-user device needs to expose those protocols.
    \item \textbf{Upstreams}: The remaining uncategorized exposures.
\end{enumerate}

\figref{fig:sankey_src_category_protocol_dst} shows a categorized representation based on the protocols and the source and destination networks.
Note that these numbers do not sum up to the values shown in \tabref{tab:non_censys_stats}, as a single exposure may belong to a different category on different providers.
A summarized table of exposures per stakeholder can be found in \tabref{tab:categorized_exposures}, which shows how different categories have their distinct characteristics.
For example, even when both user and provider exposures share the same hop count limitations, we are seeing more administrative exposures (on protocols like SSH, Telnet, and SNMP) in the provider category.

\begin{table}
	\centering
	\caption[Exposures by Stakeholder Group]{Exposures by Stakeholder. \emph{Services} shows the number of unique identifiers, \emph{Providers} the number of providers. Sorted in ascending order by (user, providers column).}
\small
\resizebox{1\columnwidth}{!}{
\begin{tabular}{lrrrrrr}
\toprule
 & \multicolumn{2}{c}{User} & \multicolumn{2}{c}{Provider} & \multicolumn{2}{c}{Upstream} \\
 \midrule
Protocol & Services & Providers & Services & Providers & Services & Providers \\
\cmidrule(lr){2-3} \cmidrule(lr){4-5}  \cmidrule(lr){6-7}

SMB & 1,064 & 29 & 0 & 0 & 171 & 14 \\
NetBIOS & 615 & 24 & 0 & 0 & 27 & 18 \\
SSH & 382 & 18 & 714 & 43 & 1,196 & 47 \\
UPnP & 107 & 18 & 0 & 0 & 13 & 9 \\
IPP & 15 & 14 & 0 & 0 & 2 & 3 \\
HTTP & 64 & 12 & 290 & 35 & 850 & 37 \\
Telnet & 39 & 9 & 133 & 31 & 304 & 42 \\
HTTPS & 12 & 4 & 105 & 31 & 586 & 40 \\
SMTP & 11 & 3 & 187 & 8 & 113 & 2 \\
SNMPv3 & 7 & 3 & 372 & 30 & 1,272 & 45 \\
FTP & 2 & 2 & 59 & 15 & 102 & 29 \\
SNMPv2 & 2 & 2 & 189 & 16 & 750 & 26 \\
\midrule
Total & 2,320 & 31 & 2,049 & 52 & 5,386 & 52 \\

\bottomrule
\end{tabular}

}
\label{tab:categorized_exposures}
\end{table}

We now discuss our findings by stakeholder.

\textbf{End Users.}
For end-user exposures, almost half of the tested providers were exposing some tested protocols;
the most common being \gls{smb} (on 29) and NetBIOS (on 24 providers).
In total numbers, these were also the most exposed services in this category, followed up with \gls{ssh} and \gls{upnp}.
As can be seen in \figref{fig:sankey_src_category_protocol_dst}, the exposed services were also predominantly found in the same aggregated RFC~1918 network,
giving some confidence that these are indeed unexpected end-user exposures.

\textbf{VPN Providers.}
In this category, the number of exposures is similarly high---and notably on more providers (52, 78~\%)---but we are seeing a rise in the number of administrative protocols such as \gls{ssh}, Telnet, and \gls{snmp}.
These protocols have also been observed for end users, but the numbers seem to clearly indicate that these exposures are more likely to come from infrastructure devices.
As noted earlier, some of these could also be from end-user devices.

\textbf{Upstreams.}
Surprisingly, the majority of exposures---also on 52 providers---belonged to this category and were deeper (in hop distance) than expected.
This seems to  indicate that the lack of network filtering is not only limited to \gls{vpn} providers.
While it is expected to see \gls{ssh} in the network infrastructure, the number of Telnet instances was likewise surprising.
As can be seen in Figure~\ref{fig:sankey_src_category_protocol_dst}, most services in this category were located in a different address space from the one assigned to our prober.

To conclude, these results reveal findings concerning \emph{all} the defined stakeholders.
We argue that \emph{almost none} of these services should be accessible for regular \gls{vpn} customers, and therefore we leave the analysis of \emph{specific vulnerabilities} out from our analyses.
Also, even if the absolute numbers of exposed services may sound small---considering how many users these services likely have---some of it can be explained by our 
opportunistic methodology. 

\subsection{Analysis of Protocol Responses}\label{sec:protocol_analyses}
In this section, we explore the exposed services more in-depth leveraging the response payloads.
We categorize the protocols into two categories: those with strong identifiers and those without such.
We start the discussion of our findings by analyzing the response payloads in the order they appear in Table~\ref{tab:non_censys_stats}.

\textbf{Strong Identifiers.}
In total, roughly 50~\% of the seen identifiers were on protocols having strong identifiers.
These were seen on 87~\% of providers and on 9~\% of the tested \gls{vpn} endpoints.
\gls{ssh} (on 84~\% of the providers), \gls{snmp}v3 (75~\%), and \gls{https} (69~\%) were more prevalent than \gls{upnp} (33~\%).

\textbf{SSH.}
During the protocol negotiation phase, \gls{ssh} servers offer a 
host key to the client.~\cite{DBLP:journals/rfc/rfc4251} 
We use the host key fingerprint as a unique identifier.
Also, as a part of the protocol negotiation, both participants send a banner with the software version that we use to categorize the server implementations~\cite{DBLP:journals/rfc/rfc4253}.

Slightly under half of the collected \gls{ssh} host keys were not seen in Censys data set (most seen being zero hops away, i.e., on the \gls{vpn} host itself),
leaving us with 2,100 (56~\% from the total) host keys from 1,200 different endpoints on over 80~\% of all tested providers.
OpenSSH was the most common implementation with 66~\% (1,400),
followed by 28~\% of \emph{dropbear}.
The most common (18~\%) OpenSSH version was 6.7p1 (from 2014).
The dropbear versions were likewise rather old, the most common being 2016.74 (41~\% of all dropbear instances) and some even older than that.

\textbf{HTTPS.}
During the \gls{tls} negotiation, a server offers its certificate allowing the client to authenticate it.
We use the \gls{tls} certificate signature hash as an identifier for \gls{https}.
In contrast to other identifiers we use in this paper, Censys~\cite{durumeric_search_2015} offers a way to query certificates from their full collection which we leverage here.
Almost half of the certificates we saw had \emph{not} been seen by \emph{any} of their scans.
We concentrate now on these 685 certificates, of which the majority (92~\%) were self-signed.

Most (80~\%) responses were positive (i.e., status 200), from which the majority (76~\%) were certified for subject name \textit{IPMI}
issued by organization \textit{Super Micro Computer} hosting a login page, implying 
that these devices were infrastructure devices rather than end-user systems.
The rest of the certificates were likewise prevalently for network equipment.

79~\% of negative responses were authentication requests 
with authentication realms hinting at
router interfaces.

\textbf{SNMPv3.}
\gls{snmp}v3 is noisy and even non-authenticated requests will yield a response from \gls{snmp}v3 ``engines'' as part of the discovery process~\cite{DBLP:journals/rfc/rfc2574}.
Contained within these responses is a so-called \textit{engine ID}, which is defined to be unique within the administrative domain (RFC3411~\cite[Section 3.1.1.1]{DBLP:journals/rfc/rfc3411}).
IETF offers a standardized format to allow automatic construction from host-specific elements (such as MAC address or \gls{ip} address),
and 93~\% of seen identifiers were using the \gls{ietf} format, with 70~\% using the MAC address format, followed by 17~\% of ``Net-SNMP Random'' format (based on the boot time and a randomly generated number~\cite{noauthor_manpage_nodate}) giving us a confidence that at least 87~\% of the identifiers are unique.

\textbf{UPnP.}
For \gls{upnp}, we use the Unique Device Name (UDN) as a unique identifier which is specified to be universally unique~\cite{upnp_arch}.
As with \gls{snmp}v3, we report only on identifiers that were seen in conjunction with other, non-Internet visible identifiers.
To our surprise, we saw merely 116 unique UDNs, half of them being Windows computers followed by 28 instances of Synology NAS devices and 14 HP iLO instances, which also explains why they were also seen deeper in the networks.

\textbf{Weak Identifiers.}
On the provider basis, \gls{http} was the most commonly exposed service with some instances being exposed on purpose.
The same may apply to \gls{smtp}, however, the rest of the protocols with weak identifiers---including Telnet, \gls{ftp}, NetBIOS, \gls{smb}, and \gls{snmp}v2 were seen on at least half of the providers---were unlikely left accessible on purpose.

\textbf{HTTP.}
For our analyses, we create a hash from the combination of the response headers (excluding headers related to cookies or caching) and the response body.
For Censys filtering, we create a less strict hash based on the following elements also available in their database: the body content, the status code, and \texttt{Server} and \texttt{WWW-Authenticate} headers.

Based on this, 3,703 (85~\%) of \gls{http} responses were not seen in Censys.
70~\% of these were caused by Squid proxy errors (containing the proxy's own \gls{ip} address, making them unique), which we ignore as proxy services that are likely exposed on purpose.
One provider confirmed that we had indeed probed its transparent proxies.

Filtering out these cases leaves 1,200 unique \gls{http} responses, of which ~50\% have a positive \gls{http} response code.
From 565 identifiers with status code 200, 53~\% were router configuration pages based on simple heuristics (e.g., \texttt{Server} header like \texttt{cisco-IOS} or a title like \texttt{pfSense -- Login}).

\textbf{Telnet.}
We received Telnet responses on 73~\% of all providers, and we use a hash over the banner as a unique identifier.
Most of the banners indicated a login screen with 95~\% containing one of ``login'', ``user'', or ``password''.
Based on banner inspection, we found a dozen ISPs whom we sent an e-mail to disclose this finding.

\textbf{NetBIOS \& SMB.}
\gls{smb} specifies a globally unique server GUID which we use as a unique identifier, giving us about 1,200 exposed services.
Out of these, only 81~\% seemed to be real GUIDs while the rest 
appeared to be device names.
As \gls{smb} responses do not contain much other information, we analyze them in combination with NetBIOS as they are often also used together on Windows systems (see Figure~\ref{fig:protocols_seen_together}).
For NetBIOS, we base our identifiers on the names (e.g., \texttt{DESKTOP-<x>} 
and the groups (e.g., \texttt{WORKGROUP}).

Half of the SMB identifiers were exposed together with NetBIOS allowing us to perform name-based grouping.
We group the identifiers using empirically collected substrings:
\begin{enumerate}
 \item Desktops (e.\,g., \texttt{DESKTOP}, \texttt{WIN-}, \texttt{MAC}),
 \item Network attached storages (e.\,g., \texttt{NAS}, \texttt{QNAP}),
 \item Servers (e.\,g., \texttt{SERVER}, \texttt{MIRROR}), and
 \item Media players (e.\,g., \texttt{KODI}).
\end{enumerate}

This covers 62~\% of all seen identifiers, the rest of the names did not have notable patterns but often names hinting towards end-user usage (e.g., \texttt{PAUL} or \texttt{LIVINGROOM}).
The most common group was desktops (44~\%), followed by around 5~\% for each of the rest of the categories.
For example, \texttt{<LOC>-MIRROR} pattern was seen for 18 geolocations (e.g., \texttt{VAN01-MIRROR}) on several providers was likely from infrastructure providers.

\textbf{SNMPv2.}
For \gls{snmp}v2, we construct our unique identifier by combining \texttt{sysDesc}, \texttt{sysName}, \texttt{sysLocation}, \texttt{sysObjectID}, and \texttt{sysContact}~\cite{DBLP:journals/rfc/rfc1213}.
When comparing to Censys data, we relaxed this filter to use only the name and the description as other information was often missing in their database.
This leaves us with 784 unique banners from 7,218 hosts.
Over 98~\% of all seen \texttt{sysName} values were unique among these hashes.
The most common names had systematic patterns containing abbreviations like \texttt{sw} (switch), \texttt{wi} (wireless), \texttt{cu} (customer), and similar hints to the infrastructure.

\textbf{Other Protocols.}
As the information gained from other protocols with weak identifiers (\gls{ftp}, \gls{smtp}, and \gls{ipp}) is rather meager and there were not that many instances of them, we concede to note that the most common \gls{ftp} implementations were PureFTPd and ProFTPd, while \texttt{exim} was the most common \gls{smtp} implementation.
Most of the \gls{smtp} instances (92~\%) had a fully qualified domain name in their banner and were mostly directly connected, implying that these exposures were likely on purpose.

\subsection{Shared Infrastructure}
\label{sec:shared_infra}

As missing network filtering may affect others besides the provider itself,
we now investigate the potential infrastructure sharing.

\textbf{Shared Entry and Vantage Points.}
The naive approach is to compare the \gls{ip} addresses of entry and vantage points among providers, much like Khan~et\,al.~\cite{DBLP:conf/imc/KhanDVSKV18} did.
Only two \gls{vpn} providers, operated by seemingly different companies, were sharing a total of 94 endpoints.
For vantage points, the situation was different: 13 providers had at least a single vantage point that was shared by another one.
Somewhat surprisingly, these 265 vantage points were not located only in some exotic countries, but on 45 different ones.

\begin{figure}
 \centering
 \includegraphics[width=\columnwidth]{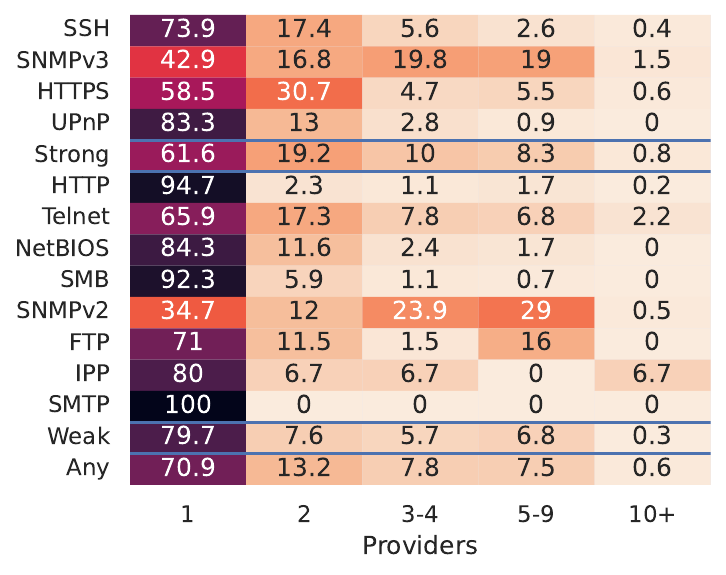}
 \caption{Number of Providers Sharing Identifiers.}
 \label{fig:shared_identifiers}
 \Description{A heatmap showing the proportion of number of providers per protocol that share identifiers.}
\end{figure}

\textbf{Shared Identifiers.}
We leverage the strong and weak identifiers
 to draw some insights about potentially shared infrastructures.
Figure~\ref{fig:shared_identifiers} shows the proportion of identifiers seen on how many providers.
The high number of shared identifiers of protocols with weak identifiers like Telnet and \gls{snmp}v2 is expected.
Almost 38~\% of strong identifiers were also seen on multiple providers: 
roughly 9~\% of \gls{ssh} host keys and 11~\% of \gls{https} certificates were seen on more than three providers.
Some examples of widely seen \gls{https} certificates were issued by ``Cisco Systems Inc.'' (common name ``nxos'', 21 providers), by ``VMware Installer'' (``localhost.localdomain'', 12 providers), and by ``MyCompany'' (``dhcp-81'', 10 providers).
The most popular \gls{ssh} host key was seen on 14 providers.

For \gls{snmp}v3, the proportion of shared identifiers is even higher, 
but closer inspection reveals that some \gls{snmp}v3 identifiers are not unique enough.
For example, the most common engine ID (\texttt{80003a8c04} reserved for ``Mikrotik''), seen on 33 providers, is not based on a MAC address.
However, a more conservative approach of accounting only MAC-address-based engine IDs, 31~\% of \gls{snmp}v3 identifiers were still seen on three or more providers.

\subsection{Responsible Disclosure}\label{sec:appendix:disclosure}
We reached out to the affected \gls{vpn} providers and informed them about our findings.
In total, we contacted 61 providers by using their preferred contact methods (i.e., e-mail, web form, bug bounty program), receiving responses from  33 providers.
We provided recommendations to fix the issues and advised them to filter the following reserved networks:
\begin{itemize}
\item Private Internets (RFC~1918~\cite{DBLP:journals/rfc/rfc1918}): \texttt{10.0.\-0.0/8}, \texttt{172.16.\-0.0/12}, and \texttt{192.168.0.0/16},
\item Link-Local (RFC~3927~\cite{DBLP:journals/rfc/rfc3927}):   \texttt{169.254.0.0/16},
\item Shared Address Space (RFC~6598~\cite{DBLP:journals/rfc/rfc6598}): \texttt{100.64.0.0/10}
\end{itemize}

Of the non-automated responses, the answers were predominantly positive, with only a single provider threatening us with potential legal actions if we were to disclose our findings.
Six providers informed us that they have put stricter filtering in place---two even awarding us a bounty---while noting that the exposed services were neither from their users nor networks under their control.
Further six providers informed us that they have escalated the issue.

We also reported the account ID numbers for the exposed cloud metadata services to Amazon, which responded swiftly that they have taken appropriate actions to secure their customers.
Furthermore, we sent e-mails to the ISPs in cases when the Telnet banners contained contact information.

\subsection{Manual Verification}
In May 2024, some time after our responsible disclosure, we wanted to check whether the situation had improved.
To this end, we tested six providers (one free and five paid) with the most end-user exposures that did not explicitly inform us about fixing the issue.
We registered an account, ran our standalone prober, and checked if a direct connection between two machines using the same configuration was possible using \gls{icmp} pings and \gls{tcp} connections.
On successful connectivity, we registered another account to verify the finding.
Overall, client-to-client communication was still possible on half of the providers tested, with two providers exposing additional unexpected hosts. 

\section{Case Studies}\label{sec:badn:casestudies}
In the following, we supplement our analyses with several case studies to highlight some interesting findings.

\textbf{Cloud Metadata Services.}
In total, 475 instances on 29 providers were responsive to our requests on \texttt{169.254.169.254}.
This data can contain some hints about the system and its configuration, information about who is administrating it (e.g., comment fields on public \gls{ssh} keys), and potentially other information.
In 26 instances on four providers, the metadata service exposed also AWS IAM credentials,
which could be used to perform actions on behalf of the linked profile.
All these credentials had an identical resource name (\texttt{instance-profile/AmazonLightsailInstanceProfile}) related to Amazon's Lightsail product.
We reported this finding to Amazon, who responded to us promptly that they have mitigated the issue.

\textbf{Exposures on Shared Address Space (CGNAT).}
Another curious finding was the responses from the Shared Address Space (\texttt{100.64/10}, RFC~6598~\cite{DBLP:journals/rfc/rfc6598}) that is reserved for \gls{cgn} implementations to avoid colliding with customer-used RFC~1918 ranges.
Although the RFC explicitly states that packets across \gls{isp} boundaries must not be forwarded, approximately 5~\% of all unique identifiers were seen responding 
from this network.
Most of these exposures were for \gls{snmp} (73~\%), but there were also instances of \gls{ssh} (13~\%), Telnet (6~\%), and \gls{https} (4~\%).

\textbf{IP Spoofing.}
473 instances on 14 providers reportedly allowed some sort of \gls{ip} spoofing.
These were predominantly on \gls{pptp} instances (465 vs. 8 with OpenVPN), highlighting likely different routing handling on endpoints.
We downloaded the session reports\footnote{\url{https://www.caida.org/projects/spoofer/screenshots.xml}} available at the website for our analyses.
All but one of the 14 providers allowed source address spoofing up to \texttt{/8} network, the largest network tested by the spoofer tool.
While this may not be surprising (e.g., Luckie~et~al. reported in 2019 that 25~\% of Spoofer project's tested autonomous systems allowed some sort of spoofing~\cite{DBLP:conf/ccs/LuckieBKKKc19}), the availability, and the lax logging policies may make these attractive for malicious users.

\section{Related Work}\label{sec:badn:related}

\textbf{Ecosystem Studies.}
The work most closely related to ours is a study by Khan~et\,al.~\cite{DBLP:conf/imc/KhanDVSKV18}, who also used several types of active network measurements to analyze 62 commercial \gls{vpn} providers and over 1,000 vantage points. Their extensive study concentrated on potential monitoring or modification, pointing out potential infrastructure sharing and usage of ``faked'' vantage point locations. 
They also provided analysis of several provider properties (e.g., marketing strategies, transparency practices, subscription models, etc.).
Ramesh~et\,al.~\cite{ramesh_vpnalyzer_2022} performed a \gls{vpn} ecosystem investigation using a cross-platform client-side tool,
where they studied 80 desktop VPNs using several types of probes to discover behavioral characteristics (like \gls{dns} features) to traffic leaks during tunnel failures.

\textbf{Information Leaks.}
Appelbaum~et\,al.~\cite{DBLP:conf/uss/RayAKF12} described a set of attack scenarios on \gls{vpn} solutions, especially concerning ``fail open'' behavior on abrupt disconnections and the lack of \gls{ipv6} routing table configuration.
Perta~et\,al.~\cite{DBLP:journals/popets/PertaBTHM15} did an experimental evaluation of \gls{dns} and \gls{ipv6} leaks on 14 commercial \gls{vpn} providers, quantifying that both problems were still prevalent in 2015.

\textbf{Client Issues.}
Ikram~et\,al.~\cite{DBLP:conf/imc/IkramVSKP16} performed a comprehensive study on the Android \gls{vpn} app ecosystem (i.e., apps with \gls{vpn} permission), including network measurement analysis on 150 apps.
They reported on several cases of different types of 
unexpected behaviors, from 3rd party trackers to malware, traffic manipulation, and even lack of encryption. 
Zhang~et\,al.~\cite{DBLP:conf/cans/ZhangLZWG17} analyzed the security of 84 OpenVPN-based Android \gls{vpn} applications, with the focus on insecure configuration settings (e.g., insecure ciphers).
Both studies report that \gls{vpn} apps for mobile phones are often OpenVPN frontends.
Bui~et\,al. analyzed the security of client configurations of 30 \gls{vpn} providers~\cite{bui_client_side_2019}.
They found insecure configurations ranging from missing server certificate validation to known pre-shared keys and credential leaks due to incorrect 
file permissions.
Recently, Xue~et\,~al.~\cite{xue_bypassing_2023} presented routing attacks that could be used to leak traffic outside the \gls{vpn} tunnel.
The premise of the described attacks lies in the fact that most VPNs rely on having routing exceptions, i.e., to make traffic to the \gls{vpn} endpoint and to the local network go outside the tunnel.

\textbf{Protocol-level Studies.}
Already in 1998, Schneier~et\,al. described several security flaws in Microsoft \gls{pptp}~\cite{DBLP:conf/ccs/SchneierM98}.
As a result, Microsoft released an improved authentication mechanism (MS-CHAPv2), which was found to be an improvement but not a complete fix, making Schneier~et\,al. to reiterate their recommendation against using \gls{pptp}~\cite{DBLP:conf/cqre/SchneierMW99}.
Horst~et\,al. have also showed how \gls{pptp} encryption can be broken in setups where RADIUS is used for authentication~\cite{foresti_breaking_2016}.
In 2018, Felsch~et\,~al.~\cite{DBLP:conf/uss/FelschGSCS18} researched \gls{ipsec}'s IKE protocols and showed how they are vulnerable to different types of vulnerabilities from dictionary attacks on low-entropy pre-shared keys to padding oracle attacks on some implementations.
Xue~et\,al.~\cite{xue_openvpn_2022} investigated how easy it would be for a network-level adversary to fingerprint and detect OpenVPN connections.
In their study, they were able to detect 85~\% of OpenVPN flows due to their specific \gls{tcp} characteristics when actively probing the server.

\section{Conclusion}
\label{sec:badn:discussion}

We performed a comprehensive empirical measurement study on popular \gls{vpn} providers to investigate unintentionally exposed services on a large scale. 
Our results show how misconfigured or missing network filtering can enable in-tunnel attackers to access otherwise inaccessible resources.
Our results reveal findings concerning \emph{all} of our defined stakeholders: end-users, \gls{vpn} providers, and their upstream providers.
The results suggest that lack of 
address filtering on private address spaces is a widespread issue,
with 40~\% of tested providers seemingly exposing end users and almost 80~\% exposing some unexpected services.

We want to emphasize that this problem is not limited to \gls{vpn} providers, but applies to all publicly accessible networked environments that leverage reserved networks for their internal routing. 
Our discussions with several \gls{vpn} providers indicate that our findings are at least partially caused by a lack of filtering by their infrastructure provider.

To help improve the situation, we release a lightweight, standalone tool that demonstrates the reactive probing approach to help network administrators test their networks for lax network configurations.
The tool and our ZGrab2 patches are available at \url{https://github.com/RUB-SysSec/TurnkeyVPNStudy/}.

\begin{acks}
We thank our anonymous reviewers for their comments that helped to improve this work.
This work was supported by the German Federal Ministry of Education and Research (BMBF) under the grant UbiTrans (16KIS1900).
\end{acks}

\printbibliography
\newpage
\appendix
\section{Tested Providers}\label{sec:appendix:providers}

Table~\ref{tab:tested_providers} lists the tested providers including the used payment method, and whether the provider was tested by Khan~et~al.~\cite{DBLP:conf/imc/KhanDVSKV18}.

\begin{center}
\tablecaption{Tested Providers\label{tab:tested_providers}}
\tablehead{%
\toprule
                Service &        IMC~\cite{DBLP:conf/imc/KhanDVSKV18} &               Payment \\
                
\midrule
}
\begin{supertabular}{lcl}
                 AirVPN & \checkmark &                 Trial \\
                Anonine & \checkmark & Prepaid CC (refunded) \\
             Anonymizer &            &         Limited Trial \\
                Astrill &            &                 Trial \\
               AzireVPN &            &               Bitcoin \\
               BolehVPN &            &               Bitcoin \\
                  BoxPN & \checkmark &            Prepaid CC \\
              BulletVPN & \checkmark &                 Trial \\
              CactusVPN &            &                 Trial \\
                   Celo & \checkmark &               Bitcoin \\
             CitizenVPN &            &                Own CC \\
            CryptoStorm &            &                Paypal \\
             CyberGhost & \checkmark &    Bitcoin (refunded) \\
             ExpressVPN & \checkmark & Prepaid CC (refunded) \\
               FinchVPN & \checkmark &            Prepaid CC \\
                FlowVPN & \checkmark &            Prepaid CC \\
             Freedom-IP & \checkmark &                Own CC \\
               FrootVPN &            &            Prepaid CC \\
              GhostPath &            &            Prepaid CC \\
              HideIPVPN & \checkmark &                 Trial \\
            HideMy.name &            &                 Trial \\
                  IBVPN & \checkmark &                 Trial \\
     Insorg / Safe-inet &            &               Bitcoin \\
               IPVanish & \checkmark &                Paypal \\
             Ironsocket & \checkmark &               Bitcoin \\
                   IVPN &            &    Bitcoin (refunded) \\
                  LeVPN & \checkmark &                Paypal \\
                LimeVPN & \checkmark &               Bitcoin \\
              LiquidVPN & \checkmark &               Bitcoin \\
                Mullvad & \checkmark &               Voucher \\
                MyIP.io & \checkmark &                Paypal \\
                NordVPN & \checkmark & Prepaid CC (refunded) \\
              OctaneVPN &            &               Bitcoin \\
               OVPN.com &            &    Bitcoin (refunded) \\
                oVPN.to &            &               Voucher \\
        Perfect Privacy &            &     Paypal (refunded) \\
Private Internet Access & \checkmark &               Bitcoin \\
          PrivateTunnel & \checkmark &                 Trial \\
             PrivateVPN & \checkmark &                 Trial \\
              ProtonVPN & \checkmark &               Bitcoin \\
                PureVPN & \checkmark &                Own CC \\
               RA4W VPN & \checkmark &            Prepaid CC \\
               SaferVPN & \checkmark &                 Trial \\
          SecureVPN.pro & \checkmark &                 Trial \\
               Seed4.Me & \checkmark &                 Trial \\
               ShadeYou & \checkmark &                 Trial \\
              SmartyDNS &            &                 Trial \\
              Surfshark &            &                Paypal \\
              SwitchVPN & \checkmark &            Prepaid CC \\
               Torguard &            &     Own CC (refunded) \\
             Trust.Zone & \checkmark &                 Trial \\
             TunnelBear & \checkmark &            Prepaid CC \\
            Unblock VPN &            &         Limited Trial \\
               VPN Gate~\cite{DBLP:conf/nsdi/NoboriS14} & \checkmark &                  Free \\
                 VPN.ac &            &               Bitcoin \\
                 VPN.ht & \checkmark &               Bitcoin \\
                VPNArea &            &               Bitcoin \\
                VPNBook & \checkmark &                  Free \\
              VPNSecure &            &               Bitcoin \\
              VPNTunnel &            &                 Trial \\
                  VPNUK & \checkmark &                 Trial \\
                VyprVPN &            &     Paypal (refunded) \\
                  Whoer &            &         Limited Trial \\
             WindScribe & \checkmark &                 Trial \\
  WiTopia / personalVPN &            &            Prepaid CC \\
               WorldVPN & \checkmark &                 Trial \\
                Zen VPN & \checkmark &                 Trial \\
\bottomrule
\end{supertabular}
\end{center}

\section{Docker Configuration}\label{sec:appendix:docker}

To allow VPN connections from the container and to simplify the data collection, we have to adjust several settings listed below. 

\noindent
\textbf{Volume bindings:}
\begin{itemize}
 \item Database access: /tmp/mongodb-27017.sock (rw)
 \item Storage: for storing auxiliary files (like PCAPs) (rw)
 \item /etc/timezone (ro)
 \item /etc/localtime (ro)
\end{itemize}

\noindent
\textbf{Exposed devices:}
\begin{itemize}
 \item For OpenVPN: /dev/net/tun
 \item For PPTP: /dev/ppp 
\end{itemize}

\noindent
\textbf{Capabilities:}
\begin{itemize}
 \item For promiscious sniffing: NET\_ADMIN
\end{itemize}

\noindent
\textbf{sysctls:}
\begin{itemize}
 \item net.ipv4.conf.all.rp\_filter = 2 (for WireGuard)
 \item net.ipv6.conf.all.disable\_ipv6 = 0 (for IPv6 OpenVPN)
\end{itemize}

\section{Protocol-specific Configurations}\label{sec:protocol_specials}
This section details protocol-specific configurations.

\textbf{OpenVPN.}
When generating configuration files from our database, we override several settings including:
\begin{itemize}
 \item \texttt{script-security} to allow executing our prober.
 \item \texttt{connect-retry}, \texttt{connect\--retry\--max}, and \texttt{connect-\\--timeout}
 (retry in 3 seconds), \texttt{connect\--retry\--max} (retry only once), \texttt{connect\--timeout} (timeout after 30 seconds) to avoid getting stuck on non-responsive endpoints.
 \item \texttt{ping-exit}: to disconnect after 15 minutes if no communication
 \item \texttt{pull-filter} to ignore server-pushed \texttt{ping\--restart} setting. For our use case it is preferred to fail than do a reconnect that would re-trigger the scanning.
\end{itemize}

We expose environment variables for the database connection using \texttt{setenv} option and hook our probing to start on \texttt{route-up} callback.
For UDP, we set \texttt{explicit\--exit\--notify} to force OpenVPN to notify the server on disconnect.
We set \texttt{auth-user-pass} to point to our generated credentials file.
On the OS level, we lower the required TLS version 
to v1 to allow probing the endpoints that do not support newer TLS versions.

\noindent
\textbf{PPTP.}
We set the following configuration parameters (among some more generic settings):
\begin{itemize}
 \item \texttt{ipparam} pointing to the instance that we can look it up from the database inside the prober
 \item \texttt{user} and \texttt{password} for authentication
 \item \texttt{require-mppe-128} and \texttt{refuse-eap} depending on the settings required by the provider.
\end{itemize}

We replace \texttt{/etc/ppp/ip-up} with a wrapper that adjusts the routing tables accordingly and launches our probing system.

\noindent
\textbf{Wireguard.}
As some of our probes are bound to a specific interface, we modify the \texttt{wg-quick} script only to use our statically defined interface.

\section{Vantage Point Locations}\label{sec:appendix:vantagepoints}
\begin{table}
	\centering
	\caption{Autonomous Systems with most Exposures.}
\begin{threeparttable}
\resizebox{\columnwidth}{!}{\begin{tabular}{lll}
\toprule
                            AS & Endpoints & Exposures \\
\midrule
NETWORK-LEAPSWITCH-IN (132335) &        15 &   1,078 \\
       WEBWERKS-AS-IN (133296) &        15 &   1,036 \\
               AS41564 (41564) &        18 &     851 \\
                   M247 (9009) &       204 &     669 \\
   ASN-QUADRANET-GLOBAL (8100) &        21 &     550 \\
              UPCLOUD (202053) &         3 &     459 \\
               AS40676 (40676) &        14 &     448 \\
            <unavailable> (-1) &       122 &     392 \\
              IDIGITAL (54643) &        10 &     362 \\
            PERFORMIVE (46562) &       744 &     330 \\
\midrule
                   291 uniques &     2,982 &  11,767 \\
\bottomrule
\end{tabular}
}

  \end{threeparttable}
\label{tab:instance_asn}
\end{table}

\begin{table}
	\centering
	\caption{Countries with most Exposures.}
\begin{threeparttable}
\resizebox{0.70\columnwidth}{!}{\begin{tabular}{lll}
\toprule
       Country & Endpoints & Exposures \\
\midrule
 United States &     1,027 &   2,506 \\
         India &        58 &   1,421 \\
        Canada &       236 &     899 \\
United Kingdom &       104 &     843 \\
       Germany &        54 &     407 \\
          None &       122 &     392 \\
        Brazil &        26 &     374 \\
   Netherlands &       167 &     297 \\
       Romania &       140 &     225 \\
        France &        61 &     194 \\
\midrule
    72 uniques &     2,983 &  10,131 \\
\bottomrule
\end{tabular}
}

  \end{threeparttable}
\label{tab:instance_country}
\end{table}

Table~\ref{tab:instance_asn} shows the most common autonomous systems and Table~\ref{tab:instance_country} the most common countries with exposed services,
highlighting that exposures are not limited to a small set of countries or autonomous systems.
The information presented in the tables is using the vantage point IP address (i.e., the egress one) for grouping.

\end{document}